# Structured Harmonic Generation via Geometric Phase Enabled Pump Shaping

**Ting-Ting Liu**[1,†], **Shi-Hui Ding**[2,†], **Chun-Yu Li**[1], **Hui Liu**[2], **Zhi-Han Zhu**[1,3,*], **Peng Chen**[2,*], **and Yan-Qing Lu**[2,*]

[1] *Wang Da-Heng Center, Heilongjiang Key Laboratory of Quantum Control, Harbin University of Science and Technology, Harbin,150080, China*
[2] *National Laboratory of Solid State Microstructures, and College of Engineering and Applied Sciences, Nanjing University, Nanjing 210093, China*
[3] *College of Quantum Science and Technology, Yanbian University, Yanji, Jilin 133002, China*
[†] *These authors contributed equally to this work.*
* *zhuzhihan@hrbust.edu.cn; chenpeng@nju.edu.cn; yqlu@nju.edu.cn*

Nonlinear optics is crucial for shaping the spatial structure of shortwave light and its interactions with matter, but achieving this through simple harmonic generation with a single pump is challenging. This study demonstrates nonlinear spin-orbit conversion using spin-dependent pump shaping via geometric phase, allowing the direct creation of desired structured harmonic waves from a Gaussian pump beam. By using the liquid-crystal flat optical elements fabricated with photoalignment, we experimentally produce higher-order cylindrically vectorial modes in second harmonic fields. We examine the vectorial spatial wavefunctions, their propagation invariance, and nonlinear spin-orbit conversion. Our results provide an efficient method for full structuring nonlinear light in broader harmonic systems, with significant applications in laser micromachining and high-energy physics.

## 1. Introduction

Optical harmonic generation involves the combination of multiple lower-frequency photons into a higher-frequency photon through parametric interactions [1,2]. This common nonlinear interaction, driven by a single pump, was the first optical nonlinear effect discovered, specifically second-harmonic generation (SHG) as observed by Franken *et al.* [3]. It has become the most prevalent technique for producing short-wavelength fields, spanning from the visible and ultraviolet spectrum to the extreme ultraviolet range [4-8]. Beyond frequency manipulation, researches on shaping the spatial structure of light via nonlinear optics also commenced with harmonic generation. Shortly after Allen *et al.* introduced the concept of optical orbital angular momentum (OAM) [9,10], their colleagues reported the conservation of OAM in the SHG of Laguerre-Gauss (LG) beams [11]. Since then, harmonic generation driven by structured light has been extensively studied across various nonlinear systems, revealing a series of unexpected phenomena, such as OAM high-harmonic generation[12,13], pulses with twist and torque[14], and harmonic transformation of spatiotemporal and topological structures[15,16].

In contrast to the well-established nonlinear techniques for frequency manipulation, shaping the complete spatial structure of light — encompassing the two-dimensional spatial spectrum and its coupling with spin dimensionality — remains highly complex for nonlinear optics [17-21]. To date, this challenge has only been addressed in a limited number of specific nonlinear interactions that were driven by multiple pumps [22,23], and remaining unachievable for harmonic generation driven by a single pump. For instance, as the scenario depicted in Fig. 1(a), following the verification of OAM conservation, early researchers discovered that the radial index of a LG pump is not conserved during harmonic generation [24]. Instead, the harmonic beam exhibits an unstable radial structure during propagation due to multi-radial-mode generation, as illustrated in Fig. 1(b). The selection rule governing the transformation of the full spatial spectrum have just recently been elucidated [25]. This situation hinders the direct generation or construction of arbitrary structured ultrashort waves or light‐matter interfaces via single-pump driven harmonic generation.

In this study, we present a vectorial nonlinear technique facilitated by the geometric phase, enabling the direct harmonic generation of arbitrarily structured light using solely a fundamental Gaussian pump. By exploiting flat optical elements based on liquid-crystal geometric phase to shape the spatial wavefunction of the applied pump, we experimentally demonstrated SHG of higher-order cylindrically vectorial (CV) modes with propagation invariance, which were polarized combination of LG conjugate pairs. The modal transformation in this nonlinear spin-orbit conversion, specifically the transformation from the polarization state of the applied pump to the spin-orbit coupled state of the resultant SHG, was investigated.

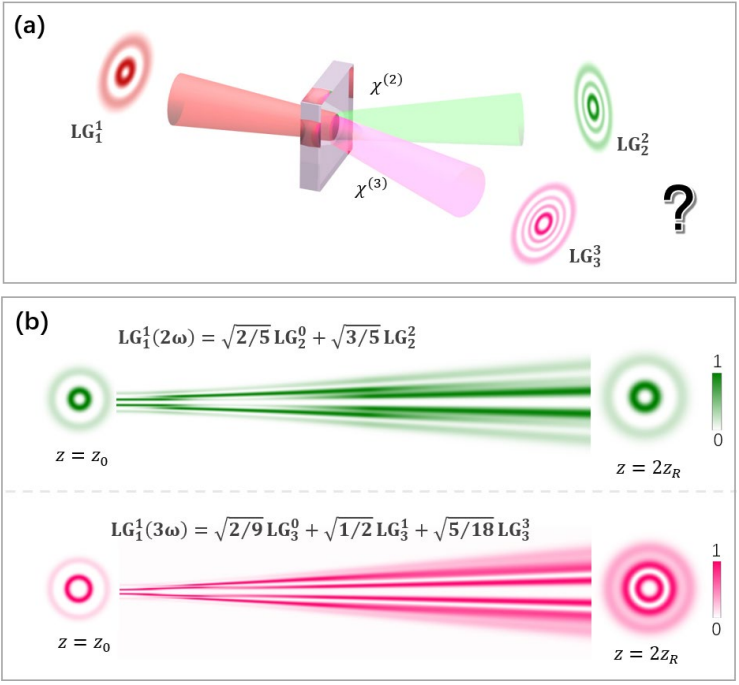

**Figure 1.** (a) Experimental Hypothesis: DO both orbital angular momentum and radial indices remain conserved during harmonic generation? (b) Actual modal compositions of second and third harmonic generations driven by a $\text{LG}^1_{+1}$ mode.

## 2. Principles and Methods

Figure 2(a) illustrates the experimental setup for implementing structured harmonic generation, which employs an SU(2) nonlinear interferometer based on a Mach-Zehnder configuration[26]. The vectorial nonlinear apparatus was designed to convert the left- and right-circular polarization components of a fundamental Gaussian pump into harmonic waves within conjugate Laguerre-Gauss modes, respectively, thereby facilitating a nonlinear spin-orbit conversion. Specifically, the input collimated Gaussian pump beam was initially modulated by a cascading quarter-wave (QWP) and half-wave plates (HWP) to achieve an arbitrarily desired polarization state, expressed as

$$\boldsymbol{E}_{in}(\omega) = \text{LG}_0^0(\vec{r})(\sqrt{a}\hat{e}_L + e^{i\theta}\sqrt{1-a}\hat{e}_R)e^{i\mathbf{k}(\omega)z}, \tag{1}$$

where $\vec{r}$ represents arbitrary transverse coordinates, $a \in [0,1]$ and $\theta$ denote the weight coefficient and intramodal phase, respectively. Then, a crucial flat-optics element, the structured geometric phase grating (SGPG)[27], for shaping pump was positioned at plane $z_0$ after cascading waveplates. This geometric phase device effectively separates two spin components of the pump into the $\pm 1^{\text{st}}$ diffraction orders and simultaneously tailors their spatial wavefunctions to $\mathscr{F}\left[\sqrt{\text{LG}^p_{+\ell}(\vec{r}, z_0)}\right]$ and its complex conjugate (c.c.), where $\ell$ and $p$ are the azimuthal and radial spatial indices of LG modes, respectively. Subsequently, two diffraction orders are focused in parallel into a vectorial nonlinear medium using a Fourier lens (L1). This medium consists of an orthogonally superposed type-I phase-matched β-Barium Borate (BBO) crystal and two orthogonal QWPs [28], facilitating simultaneous two SHGs at the Fourier plane $z_f$:

$$\begin{cases} \sqrt{a}\hat{e}_R\sqrt{LG^p_{+\ell}(\vec{r},z_f)}e^{i\mathbf{k}(\omega)z} \xrightarrow{\chi^{(2)}} a\hat{e}_L LG^p_{+\ell}(\vec{r},z_f)e^{i\mathbf{k}(2\omega)z} \\ \sqrt{1-a}\hat{e}_L\sqrt{LG^p_{-\ell}(\vec{r},z_f)}e^{i\mathbf{k}(\omega)z} \xrightarrow{\chi^{(2)}} (1-a)\hat{e}_R LG^p_{-\ell}(\vec{r},z_f)e^{i\mathbf{k}(2\omega)z} \end{cases} \quad (2)$$

Following this, two conjugate higher-order LG modes, which are already free-space eigen modes with propagation invariance, are recombined using another Fourier lens (L2) and a polarization grating (PG), resulting in a structured SHG with a CV mode

$$\boldsymbol{E}_{so}(2\omega) = \left[\sqrt{b}LG^p_{+\ell}(\vec{r})\hat{e}_L + e^{i2\theta}\sqrt{1-b}LG^p_{-\ell}(\vec{r})\hat{e}_R\right]e^{i\mathbf{k}(2\omega)z}. \quad (3)$$

where $b = a^2/[a^2 + (1-a)^2]$ is the renormalized weight coefficient. Above equations suggest that the CV mode of SHG can be all possible states within the tensor space spanned by spin-orbit bases $LG^p_{\pm\ell}(\vec{r})\hat{e}_{\pm}$, which can be geometrically represented on the surface of a unitary modal sphere [29,30]. This can be achieved experimentally through the polarization control of the applied pump using QWP and HWP. Furthermore, it has been observed that the intramodal phase is effectively doubled during nonlinear spin-orbit conversion compared to the initial pump. This phenomenon can be attributed to two primary factors: (i) the utilization of a common-path design based on geometric-phase devices, ensuring stable phase transfer within the apparatus, and (ii) the halving of the wavelength during the SHG process.

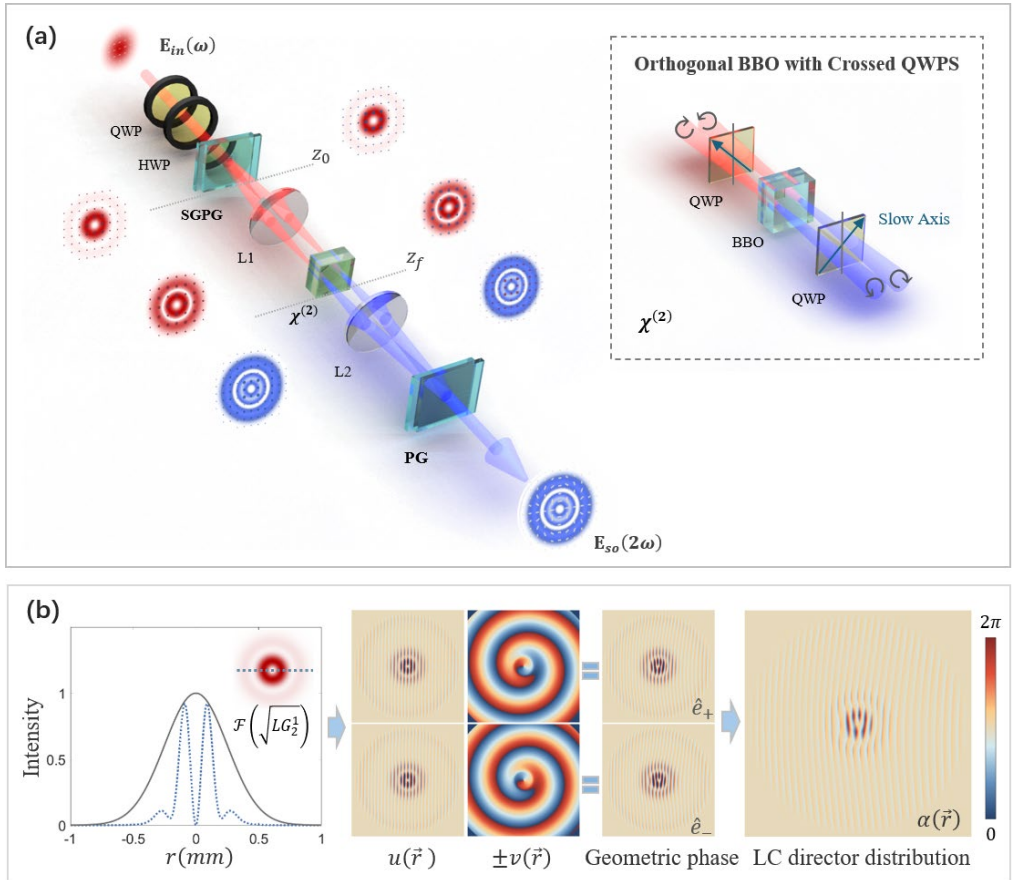

**Figure 2.** (a) Schematic of the experimental setup for the direct generation of second harmonic waves with CV modes. The key components include a half-wave plate (HWP), a quarter-wave plate (QWP), a structured geometric phase grating (SGPG), Fourier lenses with a focal length of $f = 150mm$ (L1 and L2), a polarization grating (PG), and a vector nonlinear medium ($\chi^{(2)}$) composed of two orthogonally mounted type-I β-Barium Borate (BBO) crystals and crossed QWPs, as depicted in the upper right inset. b) Principle of SGPG Design: The example illustrates the inverse design of geometric-phase conjugates and the corresponding LC-director orientations, which are capable of transforming the input Gaussian pump into $\sqrt{LG^1_{\pm 2}(\vec{r})}$ at the Fourier plane of the ±1st diffraction orders.

The foundation for achieving the aforementioned nonlinear spin-orbit conversion lies in the SGPG used for shaping pump, which was fabricated using nematic LC via the photoalignment technique[31,32]. The principle of pump shaping through an SGPG with spatially-variant LC directors, denoted as $\alpha(\vec{r})$ (orientation with respect to the *x*-axis), can be elucidated by examining its Jones matrix at the half-wave condition, given by[33,34]

$$\mathbf{M}(\vec{r}) = \begin{bmatrix} \cos2\alpha(\vec{r}) & \sin2\alpha(\vec{r}) \\ \sin2\alpha(\vec{r}) & -\cos2\alpha(\vec{r}) \end{bmatrix}, \quad (4)$$

and the resulting output field of the transmitted pump is

$$\boldsymbol{E}_{out}(\omega) = \mathbf{M}(\vec{r})\boldsymbol{E}_{in}(\omega) \\ = \mathrm{LG}_0^0(\vec{r})(\sqrt{a}\hat{e}_\mathrm{R} e^{i2\alpha(\vec{r})} + e^{i\theta}\sqrt{1-a}\hat{e}_\mathrm{L} e^{-i2\alpha(\vec{r})})e^{i\mathbf{k}(\omega)z}. \quad (5)$$

It is show that above spatially-variant polarization transformation imparts conjugate wavefronts to the two spin components, i.e., $\hat{e}_{\mathrm{L/R}} \to \hat{e}_{\mathrm{R/L}} e^{\pm i2\alpha(\vec{r})}$. Such spin-switchable wavefront modulation, in conjunction with amplitude control based on diffraction efficiency [35], tailors the transmitted pump beam into a pair of spatial-conjugate fundamental fields. More specifically, it shapes the transmitted pump beam into customed spatial structure used for the direct generation of *n*-order harmonic generations with the desired spatial wavefunction of $u(\vec{r})e^{\pm i v(\vec{r})}e^{i\mathbf{k}(n\omega)z}$. Igoring polarization components in Eq.(5), the pump shaping process can be expressed as

$$\mathrm{LG}_0^0(\vec{r},z_0)e^{\pm i2\alpha(\vec{r})}e^{i\mathbf{k}(\omega)z} \xrightarrow{\pm 1st} \mathcal{F}\left[\sqrt[n]{u(\vec{r},z_0)e^{+iv(\vec{r})}}\right]e^{i\mathbf{k}_{+1}(\omega)z} + (c.c.)e^{i\mathbf{k}_{-1}(\omega)z}, \quad (6)$$

where $\mathbf{k}_{\pm 1}(\omega)$ represents the wave vector of the transmitted pump at the $\pm 1^{\mathrm{st}}$ diffraction orders. **Figure** 2b examples the SGPG design for the direct generation of SHG ($n=2$) fields $\mathrm{LG}_{\pm 2}^1(\vec{r})e^{i\mathbf{k}(2\omega)z}$. This involves the inverse design of the geometric-phase wavefront and the corresponding LC-director orientations to tailor the spatial Fourier spectrum $\mathcal{F}\left[\sqrt{\mathrm{LG}_{\pm 2}^1(\vec{r})}\right]$ from a fundamental Gaussian beam. Notably, the design of amplitude mask has incorporated illumination correction according to the Gaussian beam profile of input pump[35].

## 3. Results and Discussions

In the proof-of-principle experiment, two pairs of conjugate LG modes $\mathrm{LG}_{\pm 2}^1(\vec{r})$ and $\mathrm{LG}_{\pm 4}^1(\vec{r})$ were selected as OAM carriers to construct the nonlinear spin-orbit conversion in structured SHG. Two crucial corresponding SGPGs were fabricated for the experiment were characterized in terms of both birefringence microstructure and beam shaping performance. **Figure** 3(a) presents the observed LC director distributions (left) and polarizing micrographs (right), with theoretical references provided in the Supplementary Materials. To assess the precision of pump beam shaping, we conducted propagation tomography of the transmitted pump beams at the $\pm 1^{\mathrm{st}}$ diffraction orders as they progressed towards the Fourier plane $z_f$, as depicted in Fig. 3(b). The observation confirms that both SGPG samples are capable of shaping the incident pump beam into the desired spatial complex amplitudes within the nonlinear crystal, i.e., $\sqrt{\mathrm{LG}_{\pm 2}^1(\vec{r})}$ and $\sqrt{\mathrm{LG}_{\pm 4}^1(\vec{r})}$.

Moreover, despite having the identical initial and Fourier patterns, the $\pm 1^{\mathrm{st}}$ diffraction orders of each sample displayed distinct variations in radial amplitude structure during propagation. This is attributed to the fact that the designed spatial complex amplitude at the $+1^{\mathrm{st}}$ order $\mathcal{F}\left[\sqrt{\mathrm{LG}_{+\ell}^p(\vec{r})}\right]$ was not a free-space eigenmode but rather a superposition state, expressed as $\sum c_p \mathrm{LG}_{+\ell/2}^p(\vec{r})$. The asynchronized accumulation of Gouy phase between successive *p* components give rise to the unstable radial structure in propagation [36]. Furthermore, the conjugate mode appearing at the $-1^{\mathrm{st}}$ order can be expressed as $\sum c_p^* \mathrm{LG}_{-\ell/2}^p(\vec{r})$, which is the 4*f* imaging of $\mathcal{F}\left[\sqrt{\mathrm{LG}_{-\ell}^p(\vec{r})}\right]$, thus exhibiting an inverse Fourier diffraction, see Supplementary Materials for details.

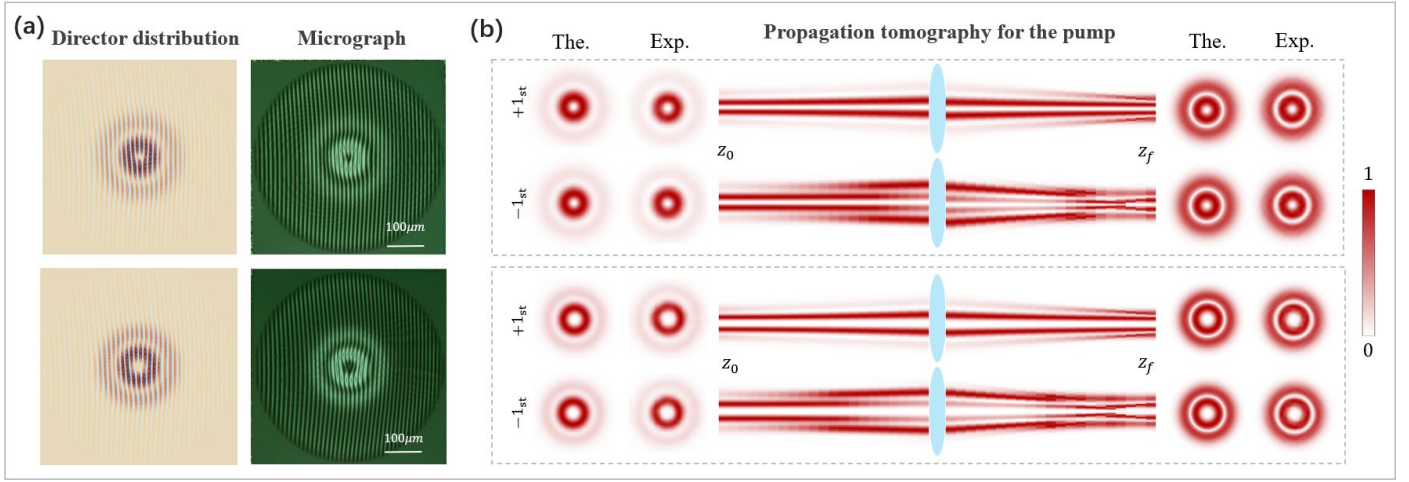

**Figure 3.** Characterizations of two fabricated SGPGs designed to shape pump spatial wavefunctions into $\sqrt{LG^1_{\pm 2}(\vec{r})}$ and $\sqrt{LG^1_{\pm 4}(\vec{r})}$ at the Fourier plane, including (a) experimental distributions of LC directors, along with polarizing micrographs captured at 0 V under crossed polarizers, and (b) propagation tomography of the tailored pump beams transmitted as they are transmitted along the $\pm 1$st diffraction orders to the Fourier plane.

Subsequently, we utilize the two previously examined SGPGs to construct the experimental setup illustrated in Fig. 2(a) and conduct the structured harmonic generation experiment. An 810 nm femtosecond pulsed laser, characterized by a 0.6 mm beam waist and horizontal polarization, was employed as the incident pump source. Accordingly, the retardance of the LC devices was adjusted to the half-wave condition for the pump wavelength by modulating their load voltage. This vectorial nonlinear apparatus facilitates pump polarization-dependent beam shaping for SHG, which, as described by Eq. (3), can produce all possible states on the spin-orbit hybrid Poincaré sphere defined by the utilized SGPG. To investigate the nonlinear spin-orbit conversion, we examine the vectorial spatial wavefunctions of the SHG that result from various polarized pump beams.

Figures 4 (a) and (b) present the experimental results of the generated SHG fields using SGPGs with OAM bases $LG^1_{\pm 2}(\vec{r})$ and $LG^1_{\pm 4}(\vec{r})$, respectively. The distinction between the two cases lies in the fact that the four states shown in Fig. 4(a) were achieved by rotating the QWP placed before the SGPG; whereas the states in Fig. 4(b) were obtained by rotating the HWP. In both cases, the observed CV modes of SHG exhibit the desired spatial distributions in amplitude, phase, and polarization, along with excellent propagation invariance, their theoretical references are provided in Supplementary Materials.

Furthermore, in contrast to the linear spin-orbit conversion, such as using a q-plate, the states on the Poincaré sphere of the resulting structured SHG were found to differ from those of the applied pump. The phenomenon aligns with the nonlinear spin-orbit conversion described by Eq. (3), which involves the transformation between the weight coefficients $a$ and $b$, as well as the doubling of the intramodal phase. This results in a nonlinear shift in the modal position on the Poincaré sphere, characterized by ellipticity and orientation $(\theta, \phi)$. To provide a clearer understanding, Fig. 4(c) depicts the ellipticity $\phi$ and orientation $\theta$ as functions of the waveplate angles $\delta(\frac{\lambda}{2})$ and $\delta(\frac{\lambda}{4})$ for both the applied pump and the generated SHG fields, see Supplementary Materials for theoretical details. The theoretical predictions are consistent with the observed state transitions via nonlinear spin-orbit conversion exampled in Figs. 4(a) and (b).

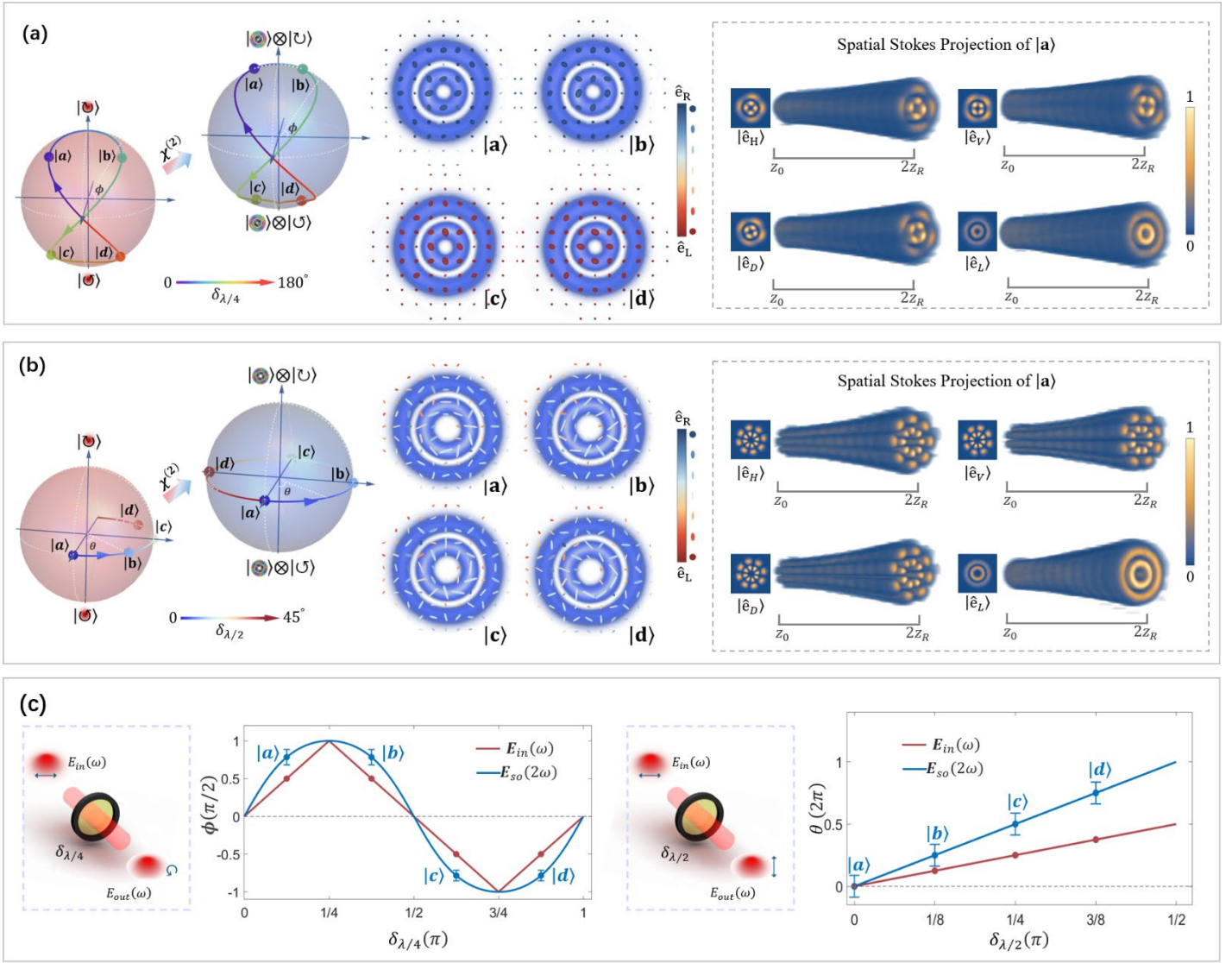

**Figure 4.** Characterizations of second-harmonic waves generated using SGPGs with OAM bases (a) $\mathrm{LG}^1_{\pm 2}(\vec{r})$ and (b) $\mathrm{LG}^1_{\pm 4}(\vec{r})$. The points on the red and bule spheres on the left illustrate the evolution of pump polarization states (red) and SHG spin-orbit states (blue) as the QWP (a) and HWP (b) are rotated. The blue patterns in the center display the vectorial beam profiles of the four SHG spin-orbit states, while the panels on the right depict the spatial Stokes propagation tomography for the SHG state |a⟩. Additional results are provided in the Supporting Information. Panel (c) shows the differential evolution of pump polarization and the associated SHG CV states with the rotation of the QWP and HWP.

## 4. Conclusion

In summary, we have experimentally demonstrated the direct generation of arbitrary structured harmonic waves through the nonlinear spin-orbit conversion of a common Gaussian pump beam. This conversion was accomplished using an SU(2) nonlinear interferometer, which enabled the transformation of two circular polarization components of a fundamental Gaussian pump into harmonic waves within desired conjugate Laguerre-Gauss modes. The pivotal element in constructing the interferometer is the SGPG capable of shaping spatial wavefunction of incident pump as required, which was fabricated using nematic LC via the photoalignment technique. We experimentally examined the vectorial spatial wavefunctions of the resultant SHGs, their propagation invariance, and corresponding relationship of state transformation in the spin-orbit conversion. This principle is also applicable to broader harmonic systems, including various higher-order harmonic generations, and thus holds significant potential for tailoring light–matter interactions in laser micromachining and high-energy physics.

## 5. Appendix

*LC Device Fabrication.*

The photoalignment agent SD1 was dissolved in dimethylformamide at a concentration of 0.35 wt%. Upon exposed to linearly polarized UV light, the SD1 molecules reorient perpendicular to the polarization direction, and then guide LC molecules to align parallel to themselves. Indium-tin-oxide glass substrates were ultrasonically bathed, UV-Ozone cleaned, spin-coated with SD1 solution, and thermally cured at 100 °C for 10 min. Two substrates were assembled and sealed with the epoxy resin mixed by 6.0 μm spacers to form a cell. The cell gap was checked by the Fabry-Pérot interference. After undergoing multi-step partly overlapping exposure by a digital-micromirror-device(DMD)-based dynamic photopatterning system, the empty LC cell was imprinted with the designed alignment patterns. Then nematic LC E7 (HCCH, China) was filled into the cell at 70 °C and gradually cooled to room temperature, yielding the electrically tunable SGPG or PG. Besides, optical and mechanical components used in the experiment were sourced from JCOPTIX.

**Funding.**

This work was supported by the National Natural Science Foundation of China (NSFC) (No. 12474324, 62222507, and 62075050), the Innovation Program for Quantum Science and Technology (No. 2021ZD0301500), the Fundamental Research Funds for the Central Universities (No. 021314380273).

**Data Availability Statement**

The data that support the findings of this study are available from the corresponding author upon reasonable request.

**Disclosures**

There are no conflicts of interest to disclose.